**Tus-*Ter*-lock immuno-PCR assays for the sensitive detection of tropomyosin-specific IgE antibodies**

Elecia B. Johnston[1], Sandip D. Kamath[2,3], Andreas L. Lopata[2,3,4] & Patrick M. Schaeffer[1,3,4]

[1]Supramolecular and Synthetic Biology Group, School of Pharmacy and Molecular Sciences, James Cook University, Douglas, QLD 4811, Australia

[2]Molecular Immunology Group, School of Pharmacy and Molecular Sciences, James Cook University, Douglas, QLD 4811, Australia

[3]Centre for Biodiscovery and Molecular Development of Therapeutics, James Cook University, Douglas, QLD 4811, Australia

[4]Comparative Genomics Centre, James Cook University, Douglas, QLD 4811, Australia

**Defined key terms**:

**Specific IgE:** Subset of IgE antibodies involved in triggering an allergic reaction to a specific protein

**Tus-*Ter*-lock:** A highly stable protein-DNA complex involved in DNA replication termination.

**Immuno-PCR:** Detection of an antigen or antibody by using a specialised protein-DNA conjugate containing a template that can be amplified and visualised by PCR.

**Tropomyosin:** An α-helical coiled-coil actin-binding protein involved in muscle contraction. This protein is the major allergenic component causing allergy to shellfish.

**ImmunoCAP:** A commercially available specific IgE quantification system for the diagnosis of allergic sensitisation.

**Abstract**

**Background:** The increasing prevalence of food allergies requires development of specific and sensitive tests capable of identifying the allergen responsible for the disease. The development of serologic tests that can detect specific IgE antibodies to allergenic proteins would therefore be highly received.

**Results:** Here we present two new quantitative immuno-PCR assays for the sensitive detection of antibodies specific to the shrimp allergen tropomyosin. Both assays are based on the self-assembling Tus-*Ter*-lock protein-DNA conjugation system. Significantly elevated levels of tropomyosin-specific IgE were detected in sera from patients allergic to shrimp.

**Conclusions:** This is the first time an allergenic protein has been fused with Tus to enable specific IgE antibody detection in human sera by quantitative immuno-PCR.
**Introduction**

Food allergies are an increasing problem world-wide, with 3-8% of people suffering some form of food allergy [1-3]. The correct diagnosis of food allergies is often cumbersome, with no single test being definitive. The diagnostic approach usually starts with a medical and dietary history followed by skin prick tests and/or serum IgE detection tests. The occurrence of false positive and false negative results with these tests means that oral food challenges often need to be completed to fully confirm clinical reactivity to a food allergen. As there are risks associated with oral food challenges and skin prick tests, accompanied with the lack of standardised procedures and reagents, the use of *in vitro* serum tests (immunoassays) for diagnosis is often preferred. Food-specific IgE assays are performed as total IgE is not well correlated with the presence of clinical allergy [4].

The first commercial test for the detection of specific IgE antibodies (sIgE) was the radioallergosorbent test, put in use in 1974 [5]. This used a radiolabelled anti-IgE to quantify the amount of IgE specific to an allergen. This was replaced by the improved fluorescence enzyme-labelled assay, ImmunoCAP (Phadia, Uppsala, Sweden) [6, 7] in 1989. The ImmunoCAP system is the most widely available commercial allergy detection test and has become the Gold standard in *in vitro* IgE quantification. This solid phase sandwich immunoassay has an allergen mix covalently coupled to a matrix, in which sIgE from serum is captured and then detected using enzyme-labelled anti-IgE antibody. Another commercial system that uses a chemiluminescent detection system has been developed to detect multiple allergens at once (MAST-CLA: multiple allergosorbent test – chemiluminescent assay; [8, 9]. This assay revealed equivalent results to ImmunoCAP [10]. In the past several other IgE detection assays were developed but none have been able to improve on the sensitivity and specificity of ImmunoCAP [11-17]. The ImmunoCAP and the Immulite systems (chemiluminescent detection system, Siemens Healthcare) [18] have a limit of quantitation of 0.1 kU/l, with the lower cut-off for allergy being 0.35 kU/l.

With the difficulty of standardising food preparations purified and recombinant allergens are being used increasingly in diagnosis. This has opened the door for the field of component-resolved diagnostics. As more allergenic proteins are characterised and pure and/or recombinant forms become available, a whole array of peptides (such as IgE-binding epitopes) can be screened for patient sensitivity. This is especially important as foods containing homologous proteins can cross-react putting an individual at further risk of allergy [19, 20]. Recently, microarray technology has been shown to be useful in the profiling of allergen sensitisation in allergic patients [21-24]. Phadia has released the ImmunoCAP ISAC (Immuno Solid-phase Allergen Chip) and it has already been put in use in some countries.

In this study we use shellfish allergy and its major allergenic protein tropomyosin (TM) to display new methods in the detection of allergen-sIgE antibodies. Shellfish allergy is among the most common of adult food allergies with approximately 2% of people being affected [25]. The identified allergens include TM [26, 27], arginine kinase [28], myosin light chain [29], and sarcoplasmic calcium-binding protein [30]. TM is an essential actin-binding protein involved in muscle contraction. Due to the highly conserved nature of tropomyosins there can be cross-reactivity against several invertebrate species of crustaceans, molluscs, insects and mites [31, 32]. New assays are therefore needed for the more sensitive detection of allergy to specific crustacean and related species.

DNA-based detection methods are being increasingly used in diagnostic assay development but they are yet to be well established in the diagnosis of allergies. The detection of a protein by immuno-PCR has become a popular method in recent years [33-35]. It is performed in a similar manner to ELISA except an antibody-DNA conjugate is used in order to amplify the signal by PCR. Common methods for coupling the antibody and DNA include a bridge with biotinylated DNA and a streptavidin-protein A fusion [36], biotinylated antibody coupled with streptavidin and biotinylated DNA [37, 38] and direct conjugation by chemical crosslinking [39]. More recently the Tus-*Ter*-lock (TT-lock) has been incorporated into a new immuno-PCR system [40-43]. Tus is a monomeric DNA-binding protein involved in DNA replication termination in *Escherichia coli* that forms a very stable complex with *Ter*-lock DNA sequences (TT-lock; $K_D$ 1 nM, half life 1 h in 0.25 M potassium chloride) [44-50]. Another DNA-based detection method replaces the PCR step with rolling circle amplification [51] and has previously been demonstrated in the detection of sIgE [52]. This method generates a concatamer of circular DNA copies attached to an anti-IgE antibody that can be detected by fluorescent complimentary oligonucleotide probes.

Here we report the development of two new quantitative immuno-PCR (qIPCR) assays building on the TT-lock qIPCR platform [40-42]. We have engineered two new detection devices for the sensitive detection of TM-specific antibodies: TM tagged with the haemagglutinin A epitope (TM-HA) and Tus-TM (Figure 1). Both assays use a *Ter*-lock

DNA extended with a single-stranded PCR template (TT-lock-T) for the detection of TM-specific antibodies by qPCR. The performances of both assays were compared using a TM-specific IgG (TM-sIgG) from rabbit. The Tus-TM TT-lock qIPCR was successfully applied for the sensitive detection of human TM-specific IgE (TM-sIgE) present in sera obtained from patients allergic to shrimp.

## Materials and Methods

**Plasmid Construction**

Cloning experiments were performed in *E. coli* DH12S. To create a Tus-tropomyosin (Tus-TM) fusion, the *Penaeus monodon* (black tiger shrimp) tropomyosin gene (*Pen m 1*) was subcloned from pRSET-A-TM (Kamath *et al*., submitted) into p-Tus-HA [42]. This was done by digesting each plasmid with *Bam*HI and *Eco*RI restriction enzymes thereby removing *Pen m 1* from pRSET-A-TM and inserting it at the 3'-end of *tus* in p-Tus-HA. This step also removed the HA-tag. The resulting plasmid was named pEJ226 and allowed the production of Tus-TM with an N-terminal hexahistidine (His6) tag (Figure 1B).

To create the TM-HA fusion protein *Pen m 1* was amplified from pRSET-A-TM (primers 5'-tgacgataaggatcgatgggg and 5'-gaattcaagcttgtagccagacagttcgctg), digested with *Bam*HI and *Hin*dIII and inserted into similarly digested pET-GFP-HA [41], thereby inserting *Pen m 1* in place of *gfp*. The resulting plasmid was named pEJ227 and allowed the production of TM-HA with a C-terminal hexahistidine tag (Figure 1A). The integrity of the inserted DNA was confirmed by DNA sequencing (Australian Genome Research Facility, Brisbane, Australia), using primers that bind to the phage T7 promoter and terminator regions in the vector.

**Protein Overproduction and Purification**

*E. coli* BL21(DE3)RIPL was used for overexpression of Tus-TM and TM-HA. Cells from a fresh transformation were used to inoculate 50-100 mL Overnight Express Instant TB Medium (Merck, Germany), supplemented with 1% glycerol (v/v), and grown at 16°C for 3-4 days. Cell pellets were resuspended in 7.5 ml/g cold lysis buffer (50 mM $NaH_2PO_4$/$Na_2HPO_4$ (pH 7.8), 300 mM NaCl, 10 % glycerol (v/v), 2 mM β-mercaptoethanol, 10 mM imidazole) and homogeneized by two passages through a French pressure cell (12,000 psi). The lysates were centrifuged at 18,000 g, at 4°C for 45 min. Soluble lysate was passed through a 2 ml nickel resin column (profinity IMAC Ni-charged resin, Bio-Rad, 156-0135) and washed with 10-15 column volumes of lysis buffer. Proteins were eluted with 3 column volumes of elution buffer (lysis buffer with 200 mM imidazole) and snap frozen for storage at -80°C.

Tus-TM required further purification. Gel filtration was carried out using a superdex 200 10/300 GL column (GE Healthcare) and a BioLogic DuoFlow system (Bio-Rad). Nickel affinity elution fractions were injected 300 µl at a time and elution proceeded at 0.25 ml/min with GF buffer (50 mM $NaH_2PO_4$/$Na_2HPO_4$ (pH 7.5), 300 mM NaCl, 10 % glycerol (v/v), 2 mM β-mercaptoethanol).

Purity of proteins was evaluated by SDS-polyacrylamide gel electrophoresis (SDS-PAGE) using 10% acrylamide gels, resolved at 200 V for ~45 min and concentrations determined by Bradford Assay.

**Electrophoretic Mobility Shift Assay (EMSA)**

Four volumes of Tus-TM (in GF buffer) or Tus-HA (in 50 mM $NaH_2PO_4$/$Na_2HPO_4$ (pH 7.8), 10 % glycerol (v/v), 2 mM β-mercaptoethanol) were mixed with one volume *TerC* [46] DNA in phosphate buffer (PB: 50 mM $NaH_2PO_4$/$Na_2HPO_4$ pH 7.8) in a 20 µl reaction volume, and incubated for 15 min at room temperature. Mixtures contained 0.5 µM *TerC* and 2 µM Tus-TM or Tus-HA. 10 µl of each reaction was run on a 2% agarose gel at 100 V for 60 min, then stained in GelRed solution for 15-30 min and visualised using a Gel Doc XR system (Bio-Rad).

**Antibodies and Patient Sera**

Antibodies used in the study were acquired as follows: goat anti-rabbit IgG conjugated to horse radish peroxidase (anti-rabbit-HRP) purchased from Promega (W4011); rabbit anti-human IgE purchased from DAKO Corporation (A0094); rat anti-HA IgG (anti-HA) purchased from Roche (11867423001); mouse anti-His$_6$ (2) purchased from Roche (04905318001); anti-rat IgG conjugated to HRP (anti-rat-HRP) purchased from Abcam (ab6734); anti-TM (rabbit anti-TM, referred to as TM-sIgG) was made in-house with the help of IMVS, SA, Australia (130 µg/ml; Kamath *et al.*, submitted).

Control serum samples were purchased from Sigma (human AB, male, H4522) and Invitrogen (human AB, 34005). Patient sera were kindly provided by Robyn E. O'Hehir and Jennifer Rolland, The Alfred Hospital, Melbourne, VIC, Australia. Subjects with a clinical history of reactivity to shellfish and one non-atopic subject were recruited by The Alfred Hospital, Allergy Clinic. Informed consent was obtained from all subjects and all experiments performed in compliance with institutional guidelines. Ethics approval for this study was granted by James Cook University Ethics committee (Project number H4313) in collaboration with The Alfred Hospital (Project number 192/07) and Monash University (MUHREC CF08/0225) Ethics Committees.

**Western blot**

TM-HA and Tus-TM were resolved by SDS-PAGE on 12% acrylamide gels and transferred to nitrocellulose membrane using a Trans-Blot Turbo blotting system (Bio-Rad). Membranes were blocked overnight at 4°C with 5% skim milk in phosphate buffered saline (PBS). All antibodies were diluted in PBS-T (PBS, 0.05% Tween 20) containing 1% skim milk. Primary antibodies TM-sIgG and anti-HA were diluted 1:2000 and 1:1000, respectively. Secondary antibodies anti-rabbit-HRP and anti-rat-HRP were diluted 1:5000 and 1:2000, respectively. Membranes were incubated for 1 h with primary antibody then washed three times for 5 min with PBS-T. Next membranes were incubated for 1 h with secondary antibody followed by three washes of 5 min with PBS-T. Bands were developed with Sigma*FAST* 3,3'-Diaminobenzidine (Sigma, D4418).

**IgE Immunoblot**

TM-HA (10 μg) was loaded on to a 12% acrylamide gel with a single well comb and resolved by SDS-PAGE. The proteins were transferred to an activated PVDF membrane (Bio-Rad). All incubation steps were done at room temperature for 1 h with shaking, unless otherwise stated. The membrane was blocked with 5% skim milk powder in PBS-T. Patient sera were diluted 1:10 in 1% skim milk-PBS-T and added to the membrane using a slot blot apparatus (Idea Scientific). Binding was performed overnight under gentle shaking at 4°C. After washing three times with PBS-T, the membrane was incubated with rabbit anti-human IgE diluted 1:8000 in 1% skim milk-PBS-T, washed three times, then incubated with anti-rabbit-HRP diluted 1:10,000 in 1% skim milk-PBS-T. After washing three times with PBS-T, the membrane was developed using enhanced chemiluminescence (ECL) [53] with Pierce ECL Western Blotting Substrate (32106).

**Preparation of anti-HA:Tus-HA:TT-lock-T detection device**

The TT-lock-T DNA probe was prepared as previously described [41]. Tus-HA was diluted to 0.5 μM in block buffer and 1.5 μl was mixed with 0.75 μl of TT-lock-T (1 μM). The mixture was incubated at room temperature for 10 min. The anti-HA (1.5 μl at 100 μg/ml) was added to the mixture and incubated for a further 5 min to obtain the anti-HA:Tus-HA:TT-lock-T detection device. The complex was then diluted in block buffer to a final concentration of 0.5 nM.

**TM-HA TT-lock qIPCR assay**

All washes were done using a Biosan microplate washer (Inteliwasher 3D-IW8, Fisher Biotec, Australia) with 200 µl wash buffer (BW: 20 mM Tris-HCl (pH 8), 150 mM NaCl, 0.005% Tween 20). All incubations were done at room temperature unless otherwise stated. Wells (MaxiSorp 96 well round-bottom plate, Nunc) were coated overnight at 4°C with 50 µl of capture antibody diluted to 10 µg/ml in PB. Wells were washed once then blocked with 100 µl block buffer (BW + 1% BSA (w/v)) for 60-90 min, and washed again. 50 µl TM-sIgG dilutions (10-fold serial dilution 10 nM - 10 fM in block buffer) were added to the wells and incubated for 90 min. Wells were then washed three times and 5 nM TM-HA (diluted in block buffer) added for 60 min. After three more washes the anti-HA:Tus-HA:TT-lock-T detection device was added and incubated for 5 min. Following five washes, 50 µl of 39/40 primer mix (39: 5'-caccgctgagcaataactagcat, 40: 5'-accgctgttgagatccagttc, diluted to 0.5 µM in water) were added and incubated for 60 min to allow the complex to dissociate.

Aliquots (10 µl) were taken from wells and mixed with 10 µl of real-time PCR mix (SensiMix SYBR and fluorescein kit, Bioline, QT615-05) to perform qPCR in 96-well plates (iCycler iQ PCR plates, Bio-Rad) with a Bio-Rad iQ5 thermocycler. PCR conditions were 95°C for 10 min, then 40 cycles of 95°C for 10 s, 55°C for 10 s, 60°C for 10 s. The 'no template controls' contained 10 µl primer mix and 10 µl PCR mix only. Positive controls contained the addition of 1 µl of detection device.

**Preparation of Tus-TM:TT-lock-T detection device**

The detection device was used at a final concentration of 0.8 nM Tus-TM and 0.1 nM TT-lock-T in the assays. TT-lock-T was first diluted to 125 nM in block buffer. Assembly of Tus-TM:TT-lock-T was performed by mixing 2 µl of Tus-TM (1 µM) and 2 µl of TT-lock-T (125 nM) together and incubating at room temperature for 10-15 min. In the 'stepwise format', the complex was diluted 1250-fold with block buffer to obtain final concentrations of 0.8 nM Tus-TM and 0.1 nM TT-lock-T respectively. In the 'mixture format', the detection device was diluted 50-fold with block buffer to obtain final concentrations of 20 nM Tus-TM and 2.5 nM TT-lock-T respectively.

**Tus-TM TT-lock qIPCR assay**

Wells were coated overnight at 4°C with 50 µl of capture antibody (anti-rabbit-HRP or anti-human IgE at 10 µg/ml in PB). Wells were washed once and blocked with 100 µl block buffer for 60-90 min.

*Stepwise format:* Serum or antibody samples were applied to wells and incubated for 90 min. Following a wash step, 50 µl of the Tus-TM:TT-lock detection device (0.8 nM Tus-TM and 0.1 nM TT-lock-T) were added to the well and incubated for 60 min. Wells were washed 5 times then 50 µl of 39/40 primer mix added and incubated for 60 min. A 10 µl aliquot of the dissociated complex was used for qPCR as for TM-HA TT-lock qIPCR assay.

*Mixture format:* 2 µl of Tus-TM:TT-lock detection device (20 nM Tus-TM and 2.5 nM TT-lock-T) were added to 50 µl of sample (as defined in the text and figure legends), mixed briefly then transferred to the wells for 90 min. Wells were washed 5 times then 50 µl of 39/40 primer mix added and incubated for 60 min. qPCR was performed as described above using a 10 µl aliquot.

**qIPCR analysis**

qPCR data was analysed using the iQ5 software (Bio-Rad, CA, USA) and graphed using GraphPad Prism (version 5.04 for Windows, GraphPad Software, CA, USA). Briefly, the fluorescence threshold baseline was manually adjusted to 180 fluorescence units and the threshold cycles (Ct) for each point extracted. ΔCt values were obtained by subtracting the experimental Ct-values from the background Ct-values generated by either the background control or negative serum control experiment. ΔCt values were plotted on a semi-log graph and where appropriate, a non-linear regression was fit to the data using the 'one site – specific binding with Hill slope' equation. The limit of detection was set at three standard deviations (SD) above the mean of the background or negative serum control samples.

## Results

### Design and characterisation of the detection devices

In this study we evaluated two different detection devices capable of detecting TM-specific antibodies that are captured on a 96-well plate. We first developed an anti-HA:Tus-HA:TT-lock-T detection device (Figure 1C) capable of binding to a TM-HA probe. This device uses the bivalency of an anti-HA IgG to link a Tus-HA:TT-lock-T complex to TM-HA. Other systems have used streptavidin as a link between components [36-38] but the biotinylation process for proteins can be laborious and the extent of biotinylation often remains incomplete, leading to problems of reproducibility. HA-tagged recombinant proteins are easy to produce and are homogeneous in nature. Using a monoclonal anti-HA antibody as the linker therefore gives reproducible control over the stoichiometry of antibody-antigen complexes.

The second detection device uses the unique combination of the Tus-TM:TT-lock-T complex (Figure 1D).

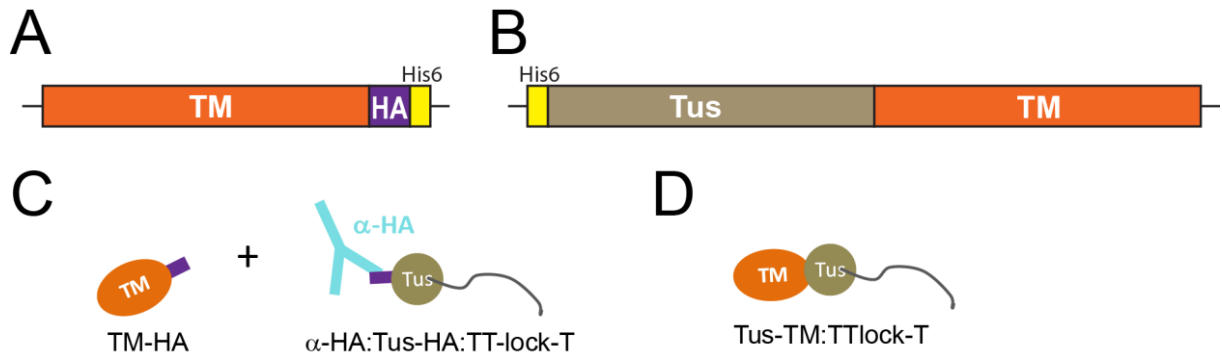

**Figure 1.** Schematic of TM detection devices. Maps of Tus-HA (A) and Tus-TM (B) gene fusions. C) Detection device for the TM-HA TT-lock qIPCR assay. The detection device consists of Tus-HA, a *Ter-lock* DNA extended with a single-stranded PCR template (TT-lock-T) and anti-HA IgG (α-HA), and is used to detect a TM-HA probe bound to TM-specific antibodies. D) Detection device for the Tus-TM TT-lock qIPCR assay. The detection device consists of Tus-TM and the TT-lock-T, and is used to detect TM-specific antibodies.

TM-HA and Tus-TM were purified by nickel affinity chromatography (Fig. 2A, B: lane NiA). The TM-HA purification was high yielding and required no further purification. Tus-TM was poorly expressed and co-eluted with several impurities, including Tus and TM fragments that were the result of proteolysis (Fig. 2B: lane NiA, Tus-TM 70 kDa, Tus & TM fragments ~40 kDa). Gel filtration was then performed to separate Tus-TM from the impurities and proteolytic fragments (Fig 2B: lane GF). Some of the Tus-TM fractions still contained a small amount of TM (from proteolysis of Tus-TM, data not shown), but the presence of these contaminants should not significantly affect the assay.

To confirm the DNA-binding activity of Tus-TM it was incubated with *TerC* [46] and visualised by EMSA. Figure 2C shows the results from an EMSA performed with Tus-TM. As a control Tus-HA:*TerC* reactions were also carried out alongside. The first lane shows *TerC*. With Tus-TM added there is a large shift in the position of the band indicating that Tus-TM is active in its *Ter* DNA-binding activity and therefore can be used in our assay.

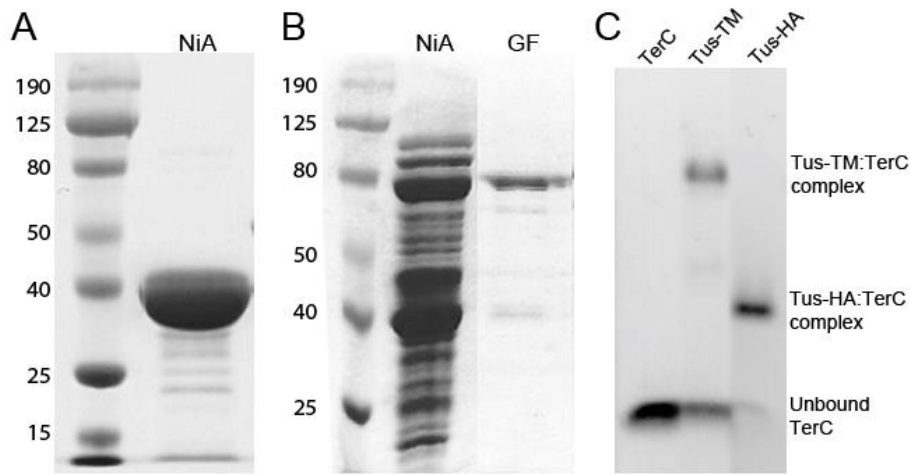

**Figure 2.** Purity of TM-HA and Tus-TM and DNA-binding activity of Tus-HA and Tus-TM. A) Purification of TM-HA by nickel affinity chromatography (NiA). B) Purification of Tus-TM by nickel affinity chromatography and gel filtration (GF). C) EMSA confirming Tus-TM and Tus-HA *Ter*-binding activity. Reactions contained 0.5 μM *TerC* and 2 μM Tus-TM or Tus-HA, 10 μl loaded. Lane 1 contains *TerC* DNA alone, lanes 2 and 3 contain Tus-TM and Tus-HA respectively.

**Development of TM-HA TT-lock qIPCR assay**

As proof-of-principle the immunoassay was first performed with known quantities of TM-sIgG obtained from rabbit. This was to simulate the capture of TM-sIgE from serum and to determine the limit of detection of the assay. For this assay plates were coated with anti-rabbit-IgG in order to capture the rabbit polyclonal TM-sIgG that was applied in a 10-fold dilution series in block buffer. The presence of TM-sIgG was then detected by the addition of TM-HA and the detection device (i.e. anti-HA:Tus-HA:TT-lock-T) and quantified by qPCR (Figure 3A).

First, the binding of both TM-sIgG and anti-HA IgG to TM-HA was demonstrated by Western Blot analysis (Figure 3B). The TM-HA TT-lock qIPCR assay was then performed with a 10-fold serial dilution of TM-sIgG (10 nM – 10 fM). A dose response curve was obtained with a detection limit of 14 pM (Figure 3C).

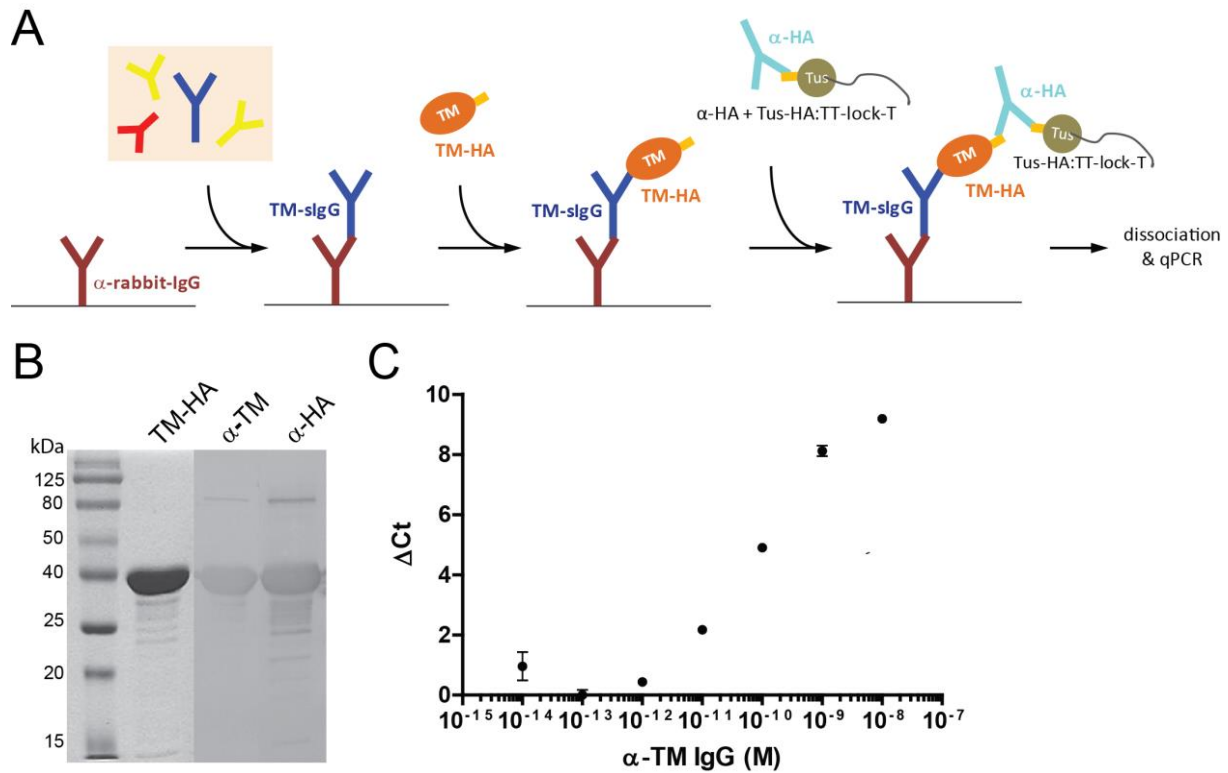

**Figure 3.** Development of TM-HA TT-lock qIPCR assay. A) Principle of the qIPCR assay for the detection of TM-sIgG. Wells were coated with anti-rabbit IgG and TM-sIgG applied in block buffer, followed by TM-HA then the anti-HA:Tus-HA:TT-lock-T detection device. B) SDS-PAGE and Western blot of TM-HA. Molecular weight marker (lane 1) and purified TM-HA (lane 2). TM-HA was probed with TM-sIgG (α-TM, lane 3) and rat anti-HA IgG (α-HA, lane 4). C) Detection of rabbit TM-sIgG by TM-HA TT-lock qIPCR assay. TM-sIgG diluted in block buffer (10 fM to 10 nM) was applied to anti-rabbit IgG-coated wells and quantifed with the detection device and qPCR. Plot of mean and SD, N=2.

As the TM-HA assay uses the bivalent properties of an antibody to bind two different proteins simultaneously, precise reagent concentrations have to be used to ensure optimal IgE detection. One downfall of this system is that it requires many incubation and wash steps, and reagents that could potentially cause reproducibility issues.

**Development of Tus-TM TT-lock qIPCR assay**

The Tus-TM TT-lock qIPCR assay was designed to reduce the number of steps and reagents required for the detection of TM-sIgE. The assay was validated with known quantities of TM-sIgG both in a stepwise and mixture format (Figure 4A). A Western Blot of Tus-TM with anti-TM and anti-His antibodies (Figure 4B) confirmed that Tus-TM contains both Tus and TM domains. The Tus-TM TT-lock qIPCR assay was first validated with TM-sIgG diluted in block buffer and then repeated with TM-sIgG diluted in neat commercial human serum. This allowed us to determine if the assay was negatively affected by the serum. No negative effects were associated with the serum. On the contrary, we noticed a reduction in background Ct-values in serum conditions compared to the same assay performed in buffer (data not shown). It is possible that some constituents of the serum saturate the surface of the well and block the non-specific binding of TT-lock-T.

Next, we tested the efficiency of the assay performed in a stepwise and in a mixture format (Figure 4A; see methods). This involved performing assays where the TM-sIgG was first applied to the wells and then the Tus-TM:TT-lock-T detection device applied in a following step after washing (stepwise); and comparing this to assays where the detection device was added directly to neat human serum (pooled commercial serum) spiked with TM-sIgG and applied in a single mixture to the well (Figure 4A). Results indicated (Figure 4C) that the mixture format performs slightly better than the stepwise format. The limit of detection for the mixture assay was calculated at 1.2 pM. Further experimental conditions were also tested (i.e. 0.4 nM Tus-TM with 50 pM or 0.4 nM TT-lock-T) in an effort to improve the limit of detection of the assay but all conditions tested yielded similar results (data not shown).

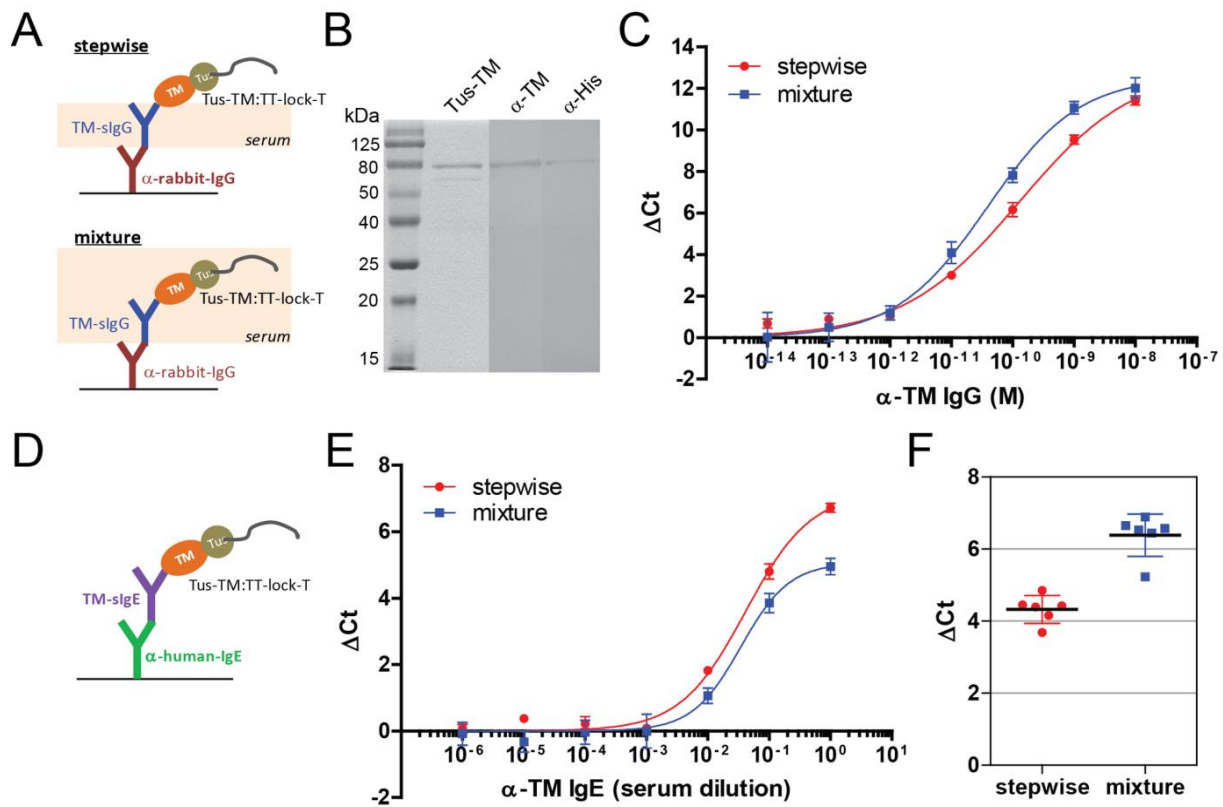

**Figure 4**. Development of Tus-TM TT-lock qIPCR assay. A) Schematic of Tus-TM sIgG detection illustrating stepwise and mixture formats. In the stepwise format, the Tus-TM:TT-lock-T detection device is added after the TM-sIgG-spiked serum was applied to anti-rabbit IgG-coated plates (α-rabbit-IgG). In the mixture format the detection device is added directly to the spiked serum. B) SDS-PAGE and Western blot of Tus-TM. Molecular weight marker (lane 1) and purified Tus-TM (lane 2). Tus-TM was probed with TM-sIgG (α-TM, lane 3) and anti-His IgG (α-His, lane 4). C) Comparison of stepwise (red) *vs* mixture (blue) formats for sIgG detection. Neat human serum was spiked with TM-sIgG (10 fM to 10 nM). Stepwise: N=2; mixture: N=2x2. D) Schematic of Tus-TM sIgE TT-lock qIPCR assay. anti-human IgE-coated plates (α-human IgE). E) Comparison of stepwise (red) *vs* mixture (blue) formats for TM-sIgE detection in serially diluted positive TM-sIgE serum (10-fold in block buffer). Stepwise: N=2; mixture: N=2x2. F) Comparison of stepwise and mixture formats for TM-sIgE detection. Positive TM-sIgE serum was diluted 1:10 in block buffer (N=6, CV=2.8%). ΔCt values were obtained by subtracting the experimental Ct-values from the background Ct-values generated by the negative control (negative patient serum diluted 1:10 in block buffer). Curves were fitted by non-linear regression: one site - specific binding with Hill slope.

To further validate and adapt our assay to the detection of TM-sIgE we coated plates with anti-human-IgE (Figure 4D). These experiments were performed using a commercial serum that strongly tested positive for TM-sIgE by immunoblot (*cf* Figure 5A, lane 4). Assays were again performed in the stepwise and mixture formats with the positive serum serially diluted in block buffer (Figure 4E). The mixture format was able to detect human sIgE in positive serum diluted up to 230-fold (Figure 4E), demonstrating the high sensitivity of this assay. Notably, the stepwise format produced a higher signal compared to the mixture format. This was found to be due to a difference in background signal between the two formats. Indeed, when both formats were performed at a 1:10 serum dilution in block buffer and Ct values obtained for the positive serum were subtracted from those of a 1:10 dilution in block buffer of a negative patient serum (confirmed negative by ImmunoCAP and Immunoblot; *cf* Figure 5A, lane1) then the mixture format systematically outperformed the stepwise format (mixture ~2 ΔCt above the stepwise format; *cf* Figure 4F). This confirmed our TM-sIgG data (Figure 3C) indicating that the mixture format would be a better choice for TM-sIgE detection.

**Patient sera testing**

A small number of patients were selected to test our new Tus-TM TT-lock qIPCR assay. Our negative control serum was obtained from a person who is not allergic to shellfish and has been confirmed negative by ImmunoCAP. This sample is designated as number 1. Two shellfish allergic patients were used who have been also tested positive by ImmunoCAP, one with a low sIgE count and one with a high count. These patients are designated 2 and 3, respectively. The final sample is from a pooled commercial lot (Invitrogen) that was found to contain a high level of TM-sIgE. This was therefore used as a positive control and designated as sample 4.

Serum samples were first tested for their reaction to recombinant TM by immunoblot against TM-HA (Figure 5A). Patients 2 and 3 (positive low and high) show a clear reaction to TM-HA with a strong band appearing near 37 kDa. This band is missing in the negative patient (1). The thick band in sample 4 demonstrates that this pooled commercial serum reacts strongly to our recombinant TM-HA. The immunoblots of sera 3 and 4 also have some other bands present that indicate a reaction to several low-abundant but antigenic *E. coli* proteins present in the sample (*cf* Figure 2A). The presence of small amounts of antigenic *E. coli* proteins in TM-HA are not concerning due to our assay design as they would be washed away during the assay and would not generate a signal even if non-specific binding were to occur. Furthermore, the Tus-TM purification included a gel filtration step reducing further the presence of these contaminants.

Next we tested the patient sera for the presence of TM-sIgE with our Tus-TM TT-lock qIPCR assay. Sera were diluted 1:10 in 50 µl reactions due to limited supply. All four sera were tested both in stepwise and mixture formats (Figure 5B). Serum from the negative patient was not reactive to TM-HA and so was chosen as a negative reference sample to determine the background signal generated by the assay. The results confirm that the mixture format performs better than the stepwise format, with the signal higher in the mixture in each case. This is most likely due to the additional wash steps in the stepwise format resulting in more complex dissociation and signal reduction. There can be a trade-off between background and sensitivity when it comes to the number of wash steps performed in an immuno-PCR, as there is with ELISA. The number of washes in a TT-lock qIPCR has previously been optimised for different assay methods [40-42] and so formed the basis of washing protocols here. At 3 SD above the negative sample the positive cut-off value is 0.9 ΔCt (mixture format). This meant that our positive low patient (2) would be borderline in the stepwise test and positive in the mixture test. As the mixture format is much more sensitive this would give us more opportunity to detect sIgE in allergic patients with low sIgE and to detect sensitised patients. All three positive patients (low ImmunoCAP positive, high ImmunoCAP positive, commercial positive) were significantly positive in this assay, with ΔCt values at 3.3, 6.4 and 6.8 respectively. With a larger patient sample size we will be able to determine a more accurate cut-off point in the future.

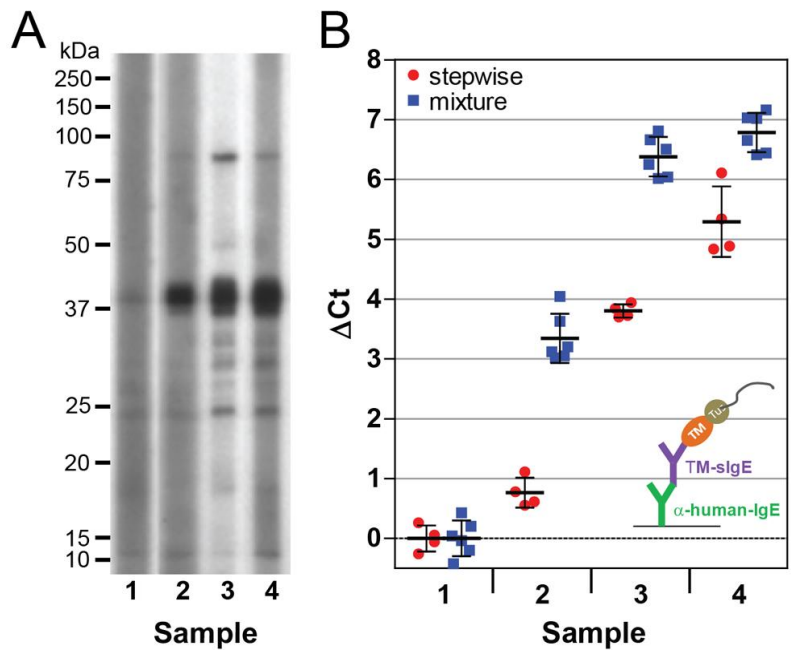

**Figure 5**. Detection of TM-sIgE in patient sera. Sample 1: negative, 2: positive (low ImmunoCAP), 3: positive (high ImmunoCAP), 4: commercial positive. A) Detection of TM-sIgE reactivity of patient sera by immunoblot. TM-HA was transferred to PVDF membrane and presence of serum TM-sIgE was visualized with anti-IgE and anti-rabbit-HRP. B) Detection of TM-sIgE in patient sera by Tus-TM TT-lock qIPCR. Reactions were performed on anti-human IgE coated plates with diluted sera (10-fold in block buffer). Baseline defined as average of negative patient Ct values. Stepwise (red): N=2x2; mixture (blue), N=3x2.

## **Discussion**

We have developed two different qIPCR assay platforms for the detection of TM-specific antibodies. Although, these assays were trialled with the major shrimp allergen TM, it is envisaged that they can be extended for use with other soluble allergens in the future. While both assays were successful in detecting TM-specific antibodies, and are of similar sensitivity, the Tus-TM method is simpler, faster and more robust. Furthermore, the TM-HA TT-lock qIPCR assay requires an additional antibody (anti-HA IgG) and fusion protein (Tus-HA) making the Tus-TM TT-lock qIPCR assay more cost-effective and the recommended choice where possible. In situations where the allergen of interest cannot be overproduced as a fusion with Tus then the HA assay system could be used as an alternative.

Other than the purchase of some oligonucleotides and an anti-human IgE, the Tus-TM sIgE TT-lock qIPCR assay only requires the production of one recombinant Tus fusion protein and the use of a quantitative thermal cycler. The assay uses only 5 µl serum per reaction which is much less than the 40 µl required for ImmunoCAP, the most commonly used commercial assay. This means it could potentially be performed with a sample from a single finger prick, a minimally invasive procedure benefiting children especially.

Using a reference TM-sIgG, the limit of detection was found to be approximately 1.2 pM, which is suitable for the detection of TM-sIgE at the lower limit cut-off for allergy. In diagnosis a positive patient is defined as having over 0.35 kU/l sIgE, equalling approximately 4.7 pM (1 kU/l = 2.4 ng/ml IgE) [54]. The sensitivity of this assay is similar to the ImmunoCAP, which has a limit of quantitation of 1.3 pM (0.1 kU/l). A recent study described an aptazyme-linked oligonucelotide IgE detection assay with a limit of detection of 1 pM [55], but this relies on the detection of total IgE as an indication of allergy. As it is known that the amount of total IgE is not correlated with the amount of allergen-specific IgE or the presence of allergy it is beneficial to develop diagnostic assays for the detection of allergen-specific IgE rather than total IgE. In this way it is also possible to test multiple allergens. With our 96-well plate assay format it would be possible to develop a screening system against multiple allergens at once. This would give the same benefits of allergy profiling that microarray offers. Correctly determining allergen sensitivity and cross-reactivity is invaluable information to the proper management of a food allergy. ImmunoCAP ISAC can give allergen-specific IgE measurements for almost 50 allergen sources or semi-quantitative measurements for 112 allergen components. Our assay provides an alternative platform for profiling which would be simpler and cheaper than microarray. In a research laboratory situation this assay would be especially attractive.

In summary, we have shown two new methods for TM-sIgE detection that could potentially be used in allergy diagnostics. The simplest of the two uses a recombinant TM tagged with Tus, a DNA binding protein, which can be measured by qPCR. The result is a detection device that will identify TM-sIgE from small amounts of serum sample. The assay was successful in detecting patients reactive to tropomyosin in a small sample group. Future studies will include a larger number of patients to further develop the diagnostic aspect of the immunoassay.

**Future Perspective**

The diagnosis of allergy is moving towards the identification of specific allergenic proteins that are relevant to each individual patient. This will lead to better control of allergies through customised management plans. The new qIPCR assays presented in this study will be very useful to achieve this goal. We expect that our new assays will be easily amenable to the detection of other soluble allergens and therefore be adopted as a routine laboratory test due to their relative simplicity.

**Executive Summary**

Background:

- Novel assays are needed that can identify specific IgE antibodies generated to allergenic proteins in sensitised patients.

Results:

- We have developed two quantitative immuno-PCR techniques that utilise a stable protein-DNA conjugate (Tus-*Ter*-lock) to detect tropomyosin-specific antibodies.

- We engineered a recombinant Tus-tropomyosin fusion and performed trial assays using a rabbit anti-tropomyosin IgG, achieving a limit of detection of 1.2 pM.

- Using our new Tus-TM TT-lock qIPCR assay, TM-specific IgE could be detected in shellfish allergic patients using only 5 µl serum per reaction.

- The technique can be easily modified for other soluble allergens or antigens to detect specific antibodies in complex media.

**Acknowledgements**

This work was supported by a grant from the Queensland Smart Futures Fund (NIRAP, Australia). AL is an ARC Future Fellow.